\documentstyle[prd,aps,preprint]{revtex}

\begin{document}

\tightenlines

\preprint{\parbox[b]{1in}{
\hbox{\tt PNUTP-00/A03}
\hbox{\tt SOGANG-HEP 276/00}}}

\draft
\title{Bosonization of QCD at high density}
\author{Deog Ki Hong$^{1,a}$, Soon-Tae Hong$^{2,b}$, and Young-Jai
Park$^{3,b}$}

\vspace{0.05in}

\address{
$^{a}$Department of Physics, Pusan National University,
Pusan 609-735, Korea
\protect\\   
$^{b}$Department of Physics,
Sogang University, Seoul 121-742, Korea
\protect\\
\vspace{0.05in}
{\footnotesize\tt $^{1}$dkhong@pnu.edu;
$^{2}$sthong, $^3$yjpark@ccs.sogang.ac.kr}}

\vspace{0.1in}

\date{\today}

\maketitle

\begin{abstract}
We describe the color-flavor locking (CFL) color superconductor
in terms of bosonic variables, where the gaped quarks are realized
as solitons, so-called superqualitons. We then show that the ground
state of the CFL color superconductor is a $Q$-matter, which is
the lowest energy state for a given fixed baryon number.
From this $Q$-matter, we calculate
the minimal energy to create a superqualiton and argue that
it is twice of the Cooper gap.
Upon quantizing the zero modes of superqualitons, we find
superqualitons have the same quantum number as the gaped quarks
and furthermore all the high spin states of superqualitons are absent
in the effective bosonic description of the CFL color superconductor.
\end{abstract}


\vfill

It is sometimes convenient to describe a system of interacting
fermions in terms of bosonic variables, since often in that
description the interaction of elementary excitations becomes
weak and perturbative approaches are applicable~\cite{Stone:1994ys}.
In this paper, we attempt to bosonize cold quark matter of three
light flavors, where the low-lying energy states are bosonic.

Due to asymptotic freedom~\cite{Gross:1973id,Politzer:1973fx},
the stable state of matter at high density will be quark
matter~\cite{Collins:1975ky}, which has been shown to exhibit
color superconductivity at low
temperature~\cite{Barrois:1977xd,Bailin:1984bm}. The color
superconducting quark matter might exist in the core of neutron
stars, since the Cooper-pair gap and the critical temperature
turn out to be quite large, of the order of $10\sim 100~{\rm
MeV}$~\cite{Rapp:1998zu,Alford:1998zt,Rapp:2000qa,Evans:1999ek,Evans:1999nf,Schafer:1999na,Son:1999uk,Hong:2000tn,Hong:2000ru,Hong:2000fh,Schafer:1999jg,Pisarski:1999nh,Pisarski:1999av,Brown:2000aq,Hsu:2000mp},
compared to the core temperature of the neutron star, which is
estimated to be $\lesssim0.7\,{\rm MeV}$~\cite{pines}.
Furthermore, it is found that, when the density is large enough
for strange quark to participate in the Cooper-pairing, not only
color symmetry but also chiral symmetry are spontaneously broken
due to so-called color-flavor locking (CFL)~\cite{Alford:1999mk}: At
low temperature, the Cooper pairs of quarks form to lock the color
and flavor indices as
\begin{eqnarray}
\left<{\psi_L}^a_{i\alpha}(\vec p){\psi_L}^b_{j\beta}(-\vec p)
\right>=-\left<{\psi_R}^a_{i\alpha}(\vec p)
{\psi_R}^b_{j\beta}(-\vec p)\right>
=\epsilon_{\alpha\beta}\epsilon^{abI}\epsilon_{ij I}\Delta (p_F),
\end{eqnarray}
where $a,b=1,2,3$ and $i,j=1,2,3$ are color and flavor indices,
respectively, and we ignore the small color sextet component in
the condensate. In this CFL phase, the
particle spectrum can be precisely mapped into that of the
hadronic phase at low density. Observing this map, Sch\"afer and
Wilczek~\cite{Schafer:1999ef,Schafer:1999pb} further conjectured
that two phases are in fact continuously connected to each other.
The CFL phase at high density is complementary to the hadronic
phase at low density. This conjecture was subsequently
supported~\cite{Hong:1999dk} by showing that quarks in the CFL
phase are realized as  Skyrmions, called superqualitons, just
like baryons are realized as Skyrmions in the hadronic phase.


Quark matter with a finite baryon number is described
by QCD with a chemical potential, which is to restrict the system
to have a fixed baryon number;
\begin{equation}
{\cal L}={\cal L}_{\rm QCD}-\mu \bar\psi_i\gamma^0\psi_i,
\end{equation}
where $\bar\psi_i\gamma^0\psi_i$ is
the quark number density and equal chemical potentials are assumed for
different flavors, for simplicity.  The ground state in the CFL phase is nothing but
the Fermi sea where all quarks are gaped by the Cooper-pairing;
the octet has a gap $\Delta$ while the singlet has $2\Delta$.
Equivalently, this system can be described in terms of bosonic
degrees of freedom, which are small fluctuations of the Cooper pairs.
Following the previous work~\cite{Hong:1999dk},
we introduce bosonic variables, defined as
\begin{equation}
{U_L}_{ai}(x)\equiv\lim_{y\to x}{\left|x-y\right|^{\gamma_m}
\over\Delta(p_F)}\,\epsilon_{abc}\epsilon_{ijk}
\psi^{bj}_{L}(-\vec v_F,x)\psi^{ck}_{L}(\vec v_F,y),
\end{equation}
where $\gamma_m$ ($\sim\alpha_s$) is the anomalous dimension of the
diquark field and $\psi(\vec v_F,x)$ denotes a quark field with
momentum close to a Fermi momentum $\mu\vec v_F$~\cite{Hong:2000tn}.
Similarly, we define $U_R$ in terms of right-handed quarks to describe
the small fluctuations of the condensate of right-handed quarks.
Since the bosonic fields, $U_{L,R}$, are colored, they will interact
with gluons. In fact, the colored massless excitations will
constitute the longitudinal components of gluons through Higgs
mechanism. Thus, the low-energy effective Lagrangian density
for the bosonic fields in the CFL phase can be written as
\begin{eqnarray}
{\cal L}_{\rm eff}=-\frac{1}{4}F_{\mu\nu}^{A}F^{\mu\nu A}
   +g_{s}G_{\mu}^{A}J^{\mu A}+
   \left[\frac{1}{4}{F}^{2}{\rm tr}(\partial_{\mu}
    U_{L}^{\dag}\partial^{\mu}U_{L})
+n_L{\cal L}_{WZW}
   +(L\leftrightarrow R)\right]+{\cal L}_m+\cdots,
\label{eff_lag}
\end{eqnarray}
where ${\cal L}_m$ is the meson mass term and the ellipsis
denotes the higher order terms in the derivative
expansion, including mixing terms between $U_L$ and $U_R$.
The gluons couple to the bosonic fields through a minimal coupling
with a conserved current, given as
\begin{equation}
J^{A\mu}={i\over 2}F^2{\rm Tr}~U_L^{-1}T^A\partial^{\mu}U_L+
{1\over 24\pi^2}\epsilon^{\mu\nu\rho\sigma}
{\rm Tr}~T^AU_L^{-1}\partial_{\nu}U_LU_L^{-1}\partial_{\rho}U_L
U_L^{-1}\partial_{\sigma}U_L+(L\leftrightarrow R)+\cdots,
\end{equation}
where the ellipsis denotes the currents from the higher order
derivative terms in Eq.~(\ref{eff_lag}).
$F$ is a quantity analogous to the pion decay constant,
calculated to be $F\sim\mu$ in the CFL color
superconductor~\cite{Son:2000cm}. The Wess-Zumino-Witten (WZW)
term~\cite{wzw} is described by the action
\begin{equation}
\Gamma_{WZW}\equiv\int{\rm d}^4x\,{\cal L}_{WZW}
=-\frac{i}{240\pi^{2}}\int_{{\sf M}}{\rm d}^{5}r\epsilon^{\mu%
\nu \alpha\beta\gamma}{\rm
tr}(l_{\mu}l_{\nu}l_{\alpha}l_{\beta}l_{\gamma})
\end{equation}
where $l_{\mu}=U_L^{\dag}\partial_{\mu}U_L$
and the integration is defined on a five-dimensional manifold
${\sf M}=V\otimes S^{1}\otimes I$ with the
three dimensional space $V$, the compactified time $S^{1}$, and
a unit interval $I$ needed for the local form of WZW term.
The coefficients of the WZW terms in the effective Lagrangian (\ref{eff_lag})
have been shown to be $n_{L,R}=1$
by matching the flavor anomalies~\cite{Hong:1999dk}, which is
later confirmed by an explicit calculation~\cite{Nowak:2000wa}.

Among the small fluctuations of condensates, the colorless
excitations correspond to genuine Nambu-Goldstone (NG) bosons,
which can be described by a color singlet combination of
$U_{L,R}$~\cite{Hong:2000ei,Casalbuoni:1999wu}, given as
\begin{equation}
\Sigma_i^j\equiv U_{Lai}U_R^{*aj}.
\end{equation}
The NG bosons transform under the $SU(3)_L\times SU(3)_R$
chiral symmetry as
\begin{equation}
\Sigma\mapsto g_L\Sigma g_R^{\dagger},\quad {\rm with}\quad
g_{L,R}\in SU(3)_{L,R}.
\end{equation}

Since the chiral symmetry is explicitly broken by the current quark
mass, the instanton effects, and the electromagnetic interaction,
the NG bosons will get mass, which has been calculated by various
groups~\cite{Son:2000cm,Hong:2000ei,Rho:2000xf,Hong:2000ng}.
Here we focus on the meson mass due to the
current strange quark mass ($m_s)$, since it will be dominant for the
intermediate density. Then, the meson mass term is simplified as
\begin{equation}
{\cal L}_m=C\, {\rm tr}(M^T\Sigma)\cdot {\rm tr} (M^*\Sigma^{\dagger})+
O(M^4),
\label{m}
\end{equation}
where $M={\rm diag}(0,0,m_s)$ and
$C\sim \Delta^4/\mu^2\,\cdot\ln(\mu^2/\Delta^2)$. (Note that in general there
will be two more mass terms quadratic in $M$. But, they all
vanish if we neglect the current mass of up and down quarks
and also the small color-sextet component of the Cooper
pair~\cite{Hong:2000ei}.)

Now, let us try to describe the CFL color superconductor in
terms of the bosonic variables. We start with the effective
Lagrangian (\ref{eff_lag}), which is good at low energy, without
putting in the quark fields. As in the Skyrme model of baryons,
we anticipate the gaped quarks
come out as solitons, made of the bosonic degrees of freedom.
That the Skyrme picture can be realized in the CFL color superconductor
is already shown in~\cite{Hong:1999dk}, but there the mass of the
soliton is not properly calculated. Here, by identifying the
correct ground state of the CFL superconductor in the bosonic
description, we find the superqualitons have same quantum numbers
as quarks with mass of the order of gap, showing that they are
really the gaped quarks in the CFL color superconductor.
Furthermore, upon quantizing the zero modes of the soliton, we find
that high spin excitations of the soliton have energy of order
of $\mu$, way beyond the scale where the effective bosonic
description is applicable, which we interpret as the absence of
high-spin quarks, in agreement with the fermionic description.
It is interesting to note that, as we will see below, by calculating
the soliton mass in the bosonic description, one finds the coupling
and the chemical potential dependence of the Cooper-pair gap,
at least numerically, which gives us a complementary way,
if not better, of estimating the gap.

As the baryon number (or the quark number) is conserved,
though spontaneously broken,~\footnote{The spontaneously
broken baryon number
just means that the states in the Fock space do not have
a well-defined baryon number. But, still the baryon number
current is conserved in the operator sense~\cite{coleman}.}
the ground state in the bosonic description should
have the same baryon (or quark) number as the ground state
in the fermionic description.
Under the $U(1)_Q$ quark number symmetry,
the bosonic fields transform as
\begin{equation}
U_{L,R}\mapsto e^{i\theta Q}U_{L,R}e^{-i\theta Q}=e^{2i\theta}U_{L,R},
\end{equation}
where $Q$ is the quark number operator,
given in the bosonic description as
\begin{equation}
Q=i\int {\rm d}^3x~{F^2\over4}
{\rm Tr}\left[U_L^{\dagger}\partial_tU_L-\partial_tU_L^{\dagger}U_L
+\left(L\leftrightarrow R\right)\right],
\end{equation}
neglecting the quark number coming from the WZW term,
since the ground state has no nontrivial topology.
The energy in the bosonic description is
\begin{equation}
E=\int{\rm d}^3x{F^2\over 4}
{\rm Tr}\left[\left|\partial_tU_L\right|^2
+\left|\vec\nabla U_L\right|^2
+\left(L\leftrightarrow R\right)\right]+E_m+\delta E,
\end{equation}
where $E_m$ is the energy due to meson mass and $\delta E$ is the
energy coming from the higher derivative terms. Assuming
the meson mass energy is positive and $E_m +\delta E\ge0$, which
is reasonable because $\Delta/F\ll1$,
we can take, dropping the positive terms due to the spatial derivative,
\begin{equation}
E\ge \int{\rm d}^3x{F^2\over 4}
{\rm Tr}\left[\left|\partial_tU_L\right|^2
+\left(L\leftrightarrow R\right)\right](\equiv E_Q).
\end{equation}
Since for any number $\alpha$
\begin{equation}
\int{\rm d}^3 x~{\rm Tr}\left[\left|U_L+\alpha i\partial_tU_L
\right|^2+\left(L\leftrightarrow R\right)\right]\ge0,
\end{equation}
we get a following Schwartz inequality,
\begin{equation}
Q^2\le  I\,E_Q,
\label{bound}
\end{equation}
where we defined
\begin{equation}
I={F^2\over 4}\int{\rm d}^3x\,{\rm Tr}\left[U_LU_L^{\dagger}
+\left(L\leftrightarrow R\right)\right].
\end{equation}
Note that the lower bound in Eq.~(\ref{bound})
is saturated for $E_Q=\omega Q$ or
\begin{equation}
U_{L,R}=e^{i\omega t} \quad{\rm with}\quad \omega={Q\over I}.
\end{equation}
The ground state of the color superconductor, which has the lowest
energy for a given quark number $Q$, is nothing but a so-called
$Q$-matter, or the interior of a very large
$Q$-ball~\cite{Coleman:1985ki,Hong:1998ur}. Since in the fermionic
description the system has the quark number $Q=\mu^3/\pi^2\int{\rm
d}^3x=\mu^3/\pi^2\cdot I/F^2$, we find, using
$F\simeq0.209\mu$~\cite{Son:2000cm},
\begin{equation}
\omega={1\over\pi^2}\left({\mu\over F}\right)^3 F
\simeq2.32\mu.
\label{fpi}
\end{equation}
By passing, we note that  $\omega$ is numerically very close to
$4\pi F$. 
The ground state of the system in the bosonic description is a
$Q$-matter whose energy per unit quark number is $\omega$.
Now, let us suppose we consider creating a $Q=1$ state out of
the ground state. In the fermionic description,
this corresponds that we excite a gaped quark in the Fermi sea into
a free state, which costs energy at least $2\Delta$.
In the bosonic description, this amounts to creating a superqualiton
out of the $Q$-matter, while reducing the quark number of the
$Q$-matter by one. Therefore, since, reducing the quark number of
the $Q$-matter by one, we gain energy $\omega$, the energy cost to
create a gaped quark from the ground state in the bosonic
description is
\begin{equation}
\delta {\cal E}=M_Q-\omega,
\label{deltae}
\end{equation}
where $M_Q$ is the energy of the superqualiton configuration.
From the relation that $2\Delta=M_Q-\omega$, later we estimate
numerically the coupling and the chemical potential dependence
of the Cooper gap.

Following the Skyrme picture of baryons in QCD at low density,
we now investigate how gaped quarks in high density QCD are realized
in its bosonic description with the Lagrangian given in
Eq.~(\ref {eff_lag})~\cite{Hong:1999dk}.
Assuming the maximal symmetry in the superqualiton,
we seek a static configuration for the field $U_{L}$ which is the
$SU(2)$ hedgehog in color-flavor in $SU(3)$
\begin{equation}
U_{Lc}(\vec{x})=\left(
\begin{array}{cc}
e^{i\vec{\tau}\cdot\hat{x}\theta (r)} & 0 \\
0 & 1
\end{array}
\right)
\label{u}
\end{equation}
where the $\tau_{i}$ ($i$=1,2,3) are Pauli matrices,
$\hat{x}\equiv\vec{x}/r$ and
$\theta (r)$ is the chiral angle determined by minimizing the static
mass $M_{0}$ given below and for unit winding number we take
$\theta (r=\infty)=0$ and $\theta (0)=\pi$.
The static configuration for the other fields are described as
\begin{equation}
U_{R}=0,~~G_{0}^{A}=\frac{x^{A}}{r}\omega (r),~~G_{i}^{A}=0.
\end{equation}

Now we consider the zero modes of the $SU(3)$ superqualiton as follows
\begin{equation}
U(\vec{x},t)={\cal A}(t)U_{Lc}(\vec{x}){\cal A}(t)^{\dag}.
\label{uxt}
\end{equation}
The Lagrangian for the zero modes is then given by
\begin{equation}
L=-M_{0}+\frac{1}{2}I_{ab}{\rm tr}({\cal A}^{\dag}\dot{\cal A}\frac{\lambda_{a}}
   {2}){\rm tr}({\cal A}^{\dag}\dot{\cal A}\frac{\lambda_{b}}{2})-\frac{i}{2}
   {\rm tr}(Y{\cal A}^{\dag}\dot{\cal A}),
\end{equation}
where $I_{ab}$ is an invariant tensor on ${\cal M}=SU(3)/U(1)$
and $Y$ is the hypercharge
\[
Y=\frac{\lambda_{8}}{\sqrt{3}}=\frac{1}{3}\left(
\begin{array}{ccc}
1 & 0 & 0\\
0 & 1 & 0\\
0 & 0 & -2\\
\end{array}
\right).
\]
Using the above static configuration, we obtain the static mass $M_{0}$ and
the tensor $I_{ab}$ as follows
\begin{eqnarray}
M_{0}&=&\frac{4\pi}{3}F^{2}\int_{0}^{\infty}{\rm d}r\left[\frac12 r^{2}\left(\frac{d\theta}{dr}\right)^{2}
    +\sin^{2}\theta+\frac{\alpha_{s}}{2\pi^{3}F^{2}}
     \left(\frac{\theta-\sin\theta\cos\theta-\pi}{2r}\right)^{2}e^{-2m_{E}r}\right]
\nonumber\\
I_{ab}&=&-\frac{32\pi}{9}F^{2}\int_{0}^{\infty}{\rm d}r
r^{2}\sin^{2}\theta = -4I_{1}
~~~~~~~~~~~~~~~~~~~~~~~~~~~~~~~~~~~(a=b=1,2,3)\nonumber\\
&=&-\frac{8\pi}{3} F^{2}\int_{0}^{\infty}{\rm d}r r^{2}(1-\cos \theta)
= -4I_{2}
~~~~~~~~~~~~~~~~~~~~~~~~~~~~~~~(a=b=4,5,6,7)\nonumber\\
&=&0
~~~~~~~~~~~~~~~~~~~~~~~~~~~~~~~~~~~~~~~~~~~~~~~~~~~~~~~~~~~~~~~~~~~~~~~~~~~~(a=b=8)
\end{eqnarray}
where $\alpha_{s}$ is the strong coupling constant and $m_{E}=\mu(6\alpha_{s}/\pi)^{1/2}$ is
the electric screening mass for the gluons.

Since ${\cal A}$ belongs to $SU(3)$, ${\cal A}^{\dag}\dot{\cal A}$ is
anti-Hermitian and traceless to be expressed as a linear combination of
$i\lambda_{a}$ as follows
\[
{\cal A}^{\dag}\dot{\cal A}=iFv^{a}\lambda_{a}=iF\left(
\begin{array}{cc}
\vec{v}\cdot\tau +\nu 1 & V \\
V^{\dag} & -2\nu\\
\end{array}
\right)
\]
where
\begin{equation}
\vec{v}=(v^{1},v^{2},v^{3}),~~V=\left(
\begin{array}{c}
v^{4}-iv^{5}\\
v^{6}-iv^{7}\\
\end{array}
\right),~~
\nu=\frac{v^{8}}{\sqrt{3}}.
\label{vs}
\end{equation}
The Lagrangian is then expressed as
\begin{equation}
L=-M_{0}+2F^{2}I_{1}\vec{v}^{2}+2F^{2}I_{2}V^{\dag}V+\frac{1}{3}NF\nu.
\end{equation}

In order to separate the SU(2) rotations from the deviations into strange
directions, we write the time-dependent rotations as follows
\[
{\cal A}(t)=\left(
\begin{array}{cc}
A(t) & 0 \\
0 & 1
\end{array}
\right)S(t) 
\]
with $A(t) \in$ SU(2) and the small rigid oscillations $S(t)$ around the
SU(2) rotations. Furthermore, in the SU(2) subgroup of SU(3), we exploit
the time-dependent collective coordinates $a^{\mu}=(a^{0},\vec{a})$
$(\mu=0,1,2,3)$ as in the SU(2) Skyrmion \cite{adkins83}
\[
A(t) = a^{0}+i\vec{a}\cdot\vec{\tau}.
\]

On the other hand the small rigid oscillations $S$, which were also used in
Ref. \cite{kleb94}, can be described as
\[
S(t)={\rm exp}(i\sum_{a=4}^{7}d^{a}\lambda_{a})={\rm exp}(i{\cal D}),
\]
where
\[
{\cal D}=\left(
\begin{array}{cc}
0 & \sqrt{2}D \\
\sqrt{2}D^{\dag} & 0
\end{array}
\right),~~ D=\frac{1}{\sqrt{2}}\left(
\begin{array}{c}
d^{4}-id^{5} \\
d^{6}-id^{7}
\end{array}
\right). 
\]

After some algebra, one can obtain the relations among the variables in
(\ref{vs}) and the SU(2) collective coordinates $a^{\mu}$ and the strange
deviations $D$ such as
\begin{eqnarray}
F\nu &=&\frac{i}{2}(D^{\dag}\dot{D}-\dot{D}^{\dag}D)-D^{\dag}(a^{0}\vec
{\dot{a}}-\dot{a}^{0}\vec{a}+\vec{a}\times\vec{\dot{a}})\cdot\vec{\tau}D
\nonumber\\
& &-\frac{i}{3}(D^{\dag}\dot{D}-\dot{D}^{\dag}D)D^{\dag}D+\cdots,
\end{eqnarray}
to yield the superqualiton Lagrangian to order $1/N$
\begin{eqnarray}
L&=&-M_{0}+2I_{1}\dot{a}^{\mu}\dot{a}^{\mu}
+4I_{2}\dot{D}^{\dag}\dot{D}+\frac{i}{6}N(D^{\dag}\dot{D}
-\dot{D}^{\dag}D)-4I_{2}m_{K}^{2}D^{\dag}D\nonumber\\
& &+2i(I_{1}-2I_{2})\{D^{\dag} (a^{0}\vec{\dot{a}}
-\dot{a}^{0}\vec{a}+\vec{a}\times\vec{\dot{a}})\cdot\vec{\tau}\dot{D}
\nonumber\\
& &-\dot{D}^{\dag}(a^{0}\vec{\dot{a}}-\dot{a}^{0}\vec{a} +\vec{a}\times\vec{%
\dot{a}})\cdot\vec{\tau}D\}
-\frac{1}{3}ND^{\dag}(a^{0}\vec{\dot{a}}
-\dot{a}^{0}\vec{a}+\vec{a}\times\vec{\dot{a}})\cdot\vec{\tau}D  \nonumber \\
& &+2\left(I_{1}-\frac{4}{3}I_{2}\right)(D^{\dag}D)(\dot{D}^{\dag}\dot{D})
-\frac{1}{2}\left(I_{1}-\frac{4}{3}I_{2}\right)(D^{\dag}\dot{D}+\dot{D}^{\dag}D)^{2}
\nonumber \\
& &+2I_{2}(D^{\dag}\dot{D}-\dot{D}^{\dag}D)^{2} -\frac{i}{9}N(D^{\dag}%
\dot{D}-\dot{D}^{\dag}D)D^{\dag}D\nonumber\\
& &+\frac{8}{3}I_{2}m_{K}^{2}(D^{\dag}D)^{2}
\label{lag}
\end{eqnarray}
where we have included the kaon mass terms proportional to the strange quark
mass which is not negligible.

The momenta $\pi^{\mu}$ and $\pi_{s}^{\alpha}$, conjugate to the
collective coordinates $a^{\mu}$ and the strange deviation
$D_{\alpha}^{\dag}$ are given by
\begin{eqnarray}
\pi^{0}&=&4I_{1}\dot{a}^{0}-2i(I_{1}-2I_{2})
(D^{\dag}\vec{a}\cdot\vec{\tau}\dot{D}-\dot{D}^{\dag}\vec{a}\cdot\vec{\tau}%
D) +\frac{1}{3}ND^{\dag}\vec{a}\cdot\vec{\tau}D  \nonumber \\
\vec{\pi}&=&4I_{1}\vec{\dot{a}}+2i(I_{1}-2I_{2})
\{D^{\dag}(a^{0}\vec{\tau}-\vec{a}\times\vec{\tau})\dot{D} -\dot{D}%
^{\dag}(a^{0}\vec{\tau}-\vec{a}\times\vec{\tau})D\}  \nonumber \\
& &-\frac{1}{3}ND^{\dag}(a^{0}\vec{\tau}-\vec{a}\times\vec{\tau})D  \nonumber \\
\pi_{s}&=&4I_{2}\dot{D}-\frac{i}{6}ND-2i(I_{1}-2I_{2})
(a^{0}\vec{\dot{a}}-\dot{a}^{0}\vec{a}+\vec{a}\times\vec{\dot{a}}) \cdot\vec{%
\tau}D  \nonumber \\
& &+2\left(I_{1}-\frac{4}{3}I_{2}\right)(D^{\dag}D)\dot{D} -\left(I_{1}-%
\frac{4}{3}I_{2}\right)(D^{\dag}\dot{D}+\dot{D}^{\dag}D)D  \nonumber \\
& &-4I_{2}(D^{\dag}\dot{D}-\dot{D}^{\dag}D)D+\frac{i}{9}N(D^{\dag}D)D
\nonumber
\end{eqnarray}
which satisfy the Poisson brackets
$$
\{a^{\mu},\pi^{\nu}\}=\delta^{\mu\nu},~~~\{D_{\alpha}^{\dag},
\pi_{s}^{\beta}\}=\{D^{\beta},\pi_{s,\alpha}^{\dag}\}=\delta_{\alpha}^{\beta}.
$$

Performing Legendre transformation, we obtain the Hamiltonian to order $1/N$
as follows
\begin{eqnarray}
H&=&M_{0}+\frac{1}{8I_{1}}\pi^{\mu}\pi^{\mu}
+\frac{1}{4I_{2}}\pi_{s}^{\dag}\pi_{s}-i\frac{N}{24I_{2}}
(D^{\dag}\pi_{s}-\pi_{s}^{\dag}D) +\left(\frac{N^{2}}{144I_{2}}\right.\nonumber\\
& &\left.+4I_{2}m_{K}^{2}\right)D^{\dag}D+i\left(\frac{1}{4I_{1}}
-\frac{1}{8I_{2}}\right)
\{D^{\dag} (a^{0}\vec{\pi}-\vec{a}\pi^{0}+\vec{a}\times\vec{\pi})
\cdot\vec{\tau}\pi_{s}  \nonumber
\\
& &-\pi_{s}^{\dag}(a^{0}\vec{\pi}-\vec{a}\pi^{0} +\vec{a}\times\vec{\pi}%
)\cdot\vec{\tau}D\} +\frac{N}{24I_{2}}D^{\dag}(a^{0}\vec{\pi}-\vec{a}%
\pi^{0} +\vec{a}\times\vec{\pi})\cdot\vec{\tau}D  \nonumber \\
& &+\left(\frac{1}{2I_{1}}-\frac{1}{3I_{2}}\right)(D^{\dag}D)
(\pi_{s}^{\dag}\pi_{s})+\left(\frac{1}{12I_{2}}-\frac{1}{8I_{1}}\right)
(D^{\dag}\pi_{s}+\pi_{s}^{\dag}D)^{2}  \nonumber \\
& &-\frac{1}{8I_{2}}\left(D^{\dag}\pi_{s}-\pi_{s}^{\dag}D\right)^{2} -i\frac{N}{24
I_{2}}(D^{\dag}\pi_{s}-\pi_{s}^{\dag}D)(D^{\dag}D)  \nonumber \\
& &+\left(\frac{N^{2}}{108I_{2}}-\frac{8}{3}I_{2}m_{K}^{2}\right)(D^{\dag}D)^{2}.
\label{hamil}
\end{eqnarray}

Applying the Batalin-Fradkin-Tyutin (BFT) scheme \cite{BFT,hong9900} to the above result, one can obtain the first class
Hamiltonian
\begin{eqnarray}
\tilde{H}&=&M_{0}+\frac{1}{8I_{1}}
(\pi^{\mu}-a^{\mu}\Phi^{2})(\pi^{\mu}-a^{\mu}\Phi^{2})\frac{a^{\nu}a^{\nu}}
{a^{\nu}a^{\nu}+2\Phi^{1}}\nonumber \\
& &+\frac{1}{4I_{2}} \pi_{s}^{\dag}\pi_{s}-i\frac{N}{24I_{2}}%
(D^{\dag}\pi_{s}-\pi_{s}^{\dag}D) +\left(\frac{N^{2}}{144I_{2}}+4I_{2}m_{K}^{2}\right)
D^{\dag}D  \nonumber \\
& &+i\left(\frac{1}{4I_{1}}-\frac{1}{8I_{2}}\right)\{D^{\dag} (a^{0}\vec{%
\pi}-\vec{a}\pi^{0}+\vec{a}\times\vec{\pi}) \cdot\vec{\tau}\pi_{s}
\nonumber \\
& &-\pi_{s}^{\dag}(a^{0}\vec{\pi}-\vec{a}\pi^{0} +\vec{a}\times\vec{\pi}%
)\cdot\vec{\tau}D\} +\frac{N}{24I_{2}}D^{\dag}(a^{0}\vec{\pi}-\vec{a}%
\pi^{0} +\vec{a}\times\vec{\pi})\cdot\vec{\tau}D  \nonumber \\
& &+\cdots
\end{eqnarray}
where the ellipsis stands for the strange-strange interaction terms of order
$1/N$ which can be readily read off from Eq. (\ref{hamil}).

Following the Klebanov and Westerberg's quantization scheme \cite{kleb94} for the
strangeness flavor direction one can obtain the Hamiltonian of the form
\begin{equation}
\tilde{H}=M_{0}+\nu  a^{\dag}a +\frac{1}
{2I_{1}} \left(\vec{I}^{2}+2c\vec{I}\cdot\vec{J}_{s}+\bar{c}\vec{J}_{s}^{2}
+\frac{1}{4}\right)  \label{hjs}
\end{equation}
where $\vec{I}$ and $\vec{J}_{s}$ are the isospin and angular momentum for
the strange quarks and
\begin{eqnarray}
\nu&=&\frac{N}{24I_{2}}(\mu_{K} -1)  \nonumber \\
c&=&1-\frac{I_{1}}{2I_{2}\mu_{K}}(\mu_{K} -1)  \nonumber \\
\bar{c}&=&1-\frac{I_{1}}{I_{2}\mu_{K}^{2}}(\mu_{K} -1)  \nonumber
\end{eqnarray}
with
$$
\mu_{K}=\left(1+\frac{m_{K}^{2}}{m_{0}^{2}}\right)^{1/2},~~~~~
m_{0}=\frac{N}{24I_{2}}.
$$
Here note that $a^{\dag}$ is creation operator for constituent strange quarks and
the factor $\frac{1}{4}$ originates from the BFT corrections
\cite{hong9900}, which are applicable to only u- and d-superqualitons.  The
Hamiltonian (\ref{hjs}) then yields the mass spectrum of superqualiton
as follows
\begin{eqnarray}
M_{Q}&=&M_{0}-(Y-\frac{1}{3})\nu+\frac{1}{2I_{1}}\left[ cJ(J+1)
+(1-c)I(I+1)\right.\nonumber\\
& &\left.+(\bar{c}-c)\frac{(Y-1/3)(Y-7/3)}{4}+\frac{1}{4}\delta_{I,1/2}\right]
\end{eqnarray}
with the total angular momentum of the quark
$\vec{J}=\vec{I}+\vec{J}_{s}$.

Unlike creating the Skyrmions out of the Dirac vacuum, in dense matter
the energy cost to create a superqualiton should be compared
with the Fermi Sea. By creating a superqualiton, we have to remove
one quark in the Fermi sea  since the total Baryon number
has to remain unchanged.  Similar to the Cooper pair mechanism \cite{cooper56},
from Eq. (\ref{deltae}),
the twice of u- and s-superqualiton masses are then given by
\begin{eqnarray}
2M_{u}&=&M_{0}+\frac{1}{2I_{1}}-\omega\nonumber\\
2M_{s}&=&M_{0}+\nu+\frac{3}{8I_{1}}\bar{c}-\omega
\end{eqnarray}
to yield the predictions for the values of $M_{u}(=M_{d})$ and $M_{s}$
\begin{equation}
\begin{array}{lll}
M_{u}=0.079\times 4\pi F, &~~~M_{s}=0.081\times 4\pi F, &~~~{\rm for}~m_{K}/F=0.1\\
M_{u}=0.079\times 4\pi F, &~~~M_{s}=0.089\times 4\pi F, &~~~{\rm for}~m_{K}/F=0.3\\
M_{u}=0.079\times 4\pi F, &~~~M_{s}=0.109\times 4\pi F, &~~~{\rm for}~m_{K}/F=0.8.
\end{array}
\end{equation}
To see if the estimated superqualiton mass is indeed the Cooper gap,
one needs to compare our numerical results with the analytic expression
for the coupling dependence of the gap.
In Table~1 we show the dependence of superqualiton
masses on the strong coupling constant $\alpha_{s}$.
By fitting the numerical results with the gap as, in the unit of
$4\pi F$,
\begin{equation}
\log(M_u)=a \log (\alpha_s)+b\alpha_s^{-1/2}+c,
\end{equation}
we get $a=0.0135$, $b=0.00341$, and $c=-2.53$.
This is very different from the analytic expression obtained in
the weak coupling limit~\cite{Son:1999uk,Hong:2000tn,Hong:2000ru,Hong:2000fh,Schafer:1999jg,Pisarski:1999nh},
\begin{equation}
\Delta\sim{\mu\over g_s^5}
\exp\left(-{3\pi^2\over\sqrt{2}g_s}\right).
\label{weak}
\end{equation}
As suggested in the reference~\cite{Rajagopal:2000rs}, the weak coupling
result (\ref{weak}) may be applicable only when the coupling is
extremely small or the chemical potential is very large.
In our numerical analysis, we are unable to probe this region.

In conclusion, we have bosonized the CFL phase of QCD at high
density, where elementary excitations are pions and kaons.
The ground state is shown to be a $Q$-matter, whose energy per
unit quark number is $2.32\mu$. The gaped quarks are realized as
solitons, so-called superqualitons. The energy to create a
superqualiton out of the ground state is argued to be
twice of the gap, which is checked numerically by calculating
the superqualiton mass as the coupling changes.
Finally, we have quantized the zero modes of superqualiton
and find that the mass of high-spin states is larger than
the chemical potential, which is interpreted as an
absence of such states in the bosonized theory, in agreement
with the fermionic description.

\acknowledgments
We are grateful to Krishna Rajagopal for a useful conversation
and to Y. J. Sohn for the help in the numerical analysis.
DKH wishes to thank the Aspen Center for Physics for its kind
hospitality during the completion of this work.
The work of DKH was supported in part by the KOSEF
through the Korean-USA  Cooperative Science Program, 1999, and
also by KOSEF grant number 1999-2-111-005-5.
STH and YJP acknowledge financial support from the Korea Research
Foundation grant number KRF-2000-015-DP0070.

\vfil

\newpage
\begin{table}[t]
\caption{The dependence of qualiton masses on the coupling $\alpha_{s}$ with $m_{K}/F=0.3$}
\begin{center}
\begin{tabular}{crrrr}
$\alpha_{s}$  &$M_{Q}(u)/4\pi F$  &$M_{Q}(s)/4\pi F$
              &$M_{u}/4\pi F$     &$M_{s}/4\pi F$\\
\hline
     0.050    &1.040    &1.061    &0.078    &0.089\\
     0.100    &1.040    &1.061    &0.078    &0.089\\
     0.150    &1.041    &1.061    &0.079    &0.089\\
     0.200    &1.041    &1.061    &0.079    &0.089\\
     0.250    &1.041    &1.061    &0.079    &0.089\\
     0.300    &1.041    &1.062    &0.079    &0.089\\
     0.350    &1.041    &1.062    &0.079    &0.089\\
     0.400    &1.042    &1.062    &0.079    &0.089\\
     0.450    &1.042    &1.062    &0.079    &0.089\\
     0.500    &1.042    &1.062    &0.079    &0.089\\
     0.550    &1.042    &1.062    &0.079    &0.089\\
     0.600    &1.042    &1.062    &0.079    &0.089\\
     0.650    &1.042    &1.062    &0.079    &0.090\\
     0.700    &1.042    &1.063    &0.079    &0.090\\
     0.750    &1.042    &1.062    &0.079    &0.090\\
     0.800    &1.042    &1.063    &0.079    &0.090\\
     0.850    &1.042    &1.063    &0.079    &0.090\\
     0.900    &1.042    &1.063    &0.079    &0.090\\
     0.950    &1.042    &1.063    &0.080    &0.090\\
     1.000    &1.043    &1.063    &0.080    &0.090\\
\end{tabular}
\end{center}
\end{table}


\end{document}